\documentclass[aps,prl,floatfix,twocolumn,superscriptaddress,showpacs]{revtex4}

\newcommand{\be}{\begin{equation}}
\newcommand{\ee}{\end{equation}}
\newcommand{\bea}{\begin{eqnarray}}
\newcommand{\eea}{\end{eqnarray}}

\usepackage{bm}
\usepackage{amssymb}
\usepackage[dvips]{graphicx}

\begin{document}

\author{M. Nizama}
\affiliation{Centro At\'omico Bariloche and Instituto Balseiro, Comisi\'on\\
Nacional de Energ\'{\i}a At\'omica, \\
8400 Bariloche, Argentina\\}
\author{ K. Hallberg}
\affiliation{Centro At\'omico Bariloche and Instituto Balseiro, Comisi\'on\\
Nacional de Energ\'{\i}a At\'omica, \\
8400 Bariloche, Argentina\\}
\author{J. d'Albuquerque e Castro}
\affiliation{Instituto de Fisica, Universidade Federal do Rio de Janeiro, RJ, Brazil}
\title{Static and dynamical properties of elliptic quantum corrals}

\begin{abstract}
Due to their focalizing properties, elliptic quantum corrals present very
interesting physical behaviours. In this work we analyze static and
dynamical properties of these systems. Results are presented for realistic values of 
the parameters which might be useful for comparison with experiments.
We study non-interacting corrals and their
response to several kinds of external perturbations, observing that, for
realistic values of the Fermi level, the dynamics involves only a few number
of excited states, making the system quite robust with respect to
possible sources of decoherence. We also study the system in the presence of
two $S=1/2$ impurities located at its foci which interact with the electrons
of the ellipse via a superexchange interaction $J$. This system is
diagonalized numerically and properties such as the spin gap and spin-spin
static and dynamical correlations are studied. We find that, for small $J$,
both spins are locked in a singlet or triplet state, for even or odd filling
respectively, and its spin dynamics consists mainly of a single peak above
the spin gap. In this limit we can define an effective Heisenberg
Hamiltonian to describe the low-energy properties. For larger $J$, more
states are involved and the localized spins decorrelate, in a similar manner
as the RKKY-Kondo transition for the two-impurity problem.
\end{abstract}
\maketitle




\section{Introduction}

The recent advancement in the knowledge and fabrication techniques of
nanoscopic systems has revealed a great variety of novel physical
behaviours, a striking example of which is the observation of the
phenomenon now known as the {\it quantum mirage} \cite{manoharan}. This
effect arises in nanoscopic systems called quantum corrals, where the
electron wave functions are confined to a finite region by means of a
barrier formed by atoms encircling a closed region in space.

Quantum corrals are built by positioning atoms, typically transition metal
atoms, along a closed line on the clean surfaces of noble metals. In their
recent experiments, Manoharan et. al.\cite{manoharan} have built elliptical
corrals with Co atoms on the (111) surface of Cu. The Cu (111) surface has a
band of surface states, orthogonal to the bulk states, which can be
represented as a two dimensional electron gas confined at the surface. The
Fermi level is placed at 450 meV above the bottom of the surface state band.
The atoms forming the corral act as scattering centers which tend to confine
the surface electrons inside the corral.

A number of theoretical papers analyzed these experiments with one or more
impurities considering different configurations \cite%
{agam,aligia,weissmann,porras,fiete,armandochiappe,lobos,nos1,nos2} (see
also complete review articles in \cite{armandoreview,fiete1}). In refs. \cite%
{nos1,nos2} it was suggested that, as a consequence of the focalizing
properties of quantum elliptic corrals, two impurities located at the foci
of the system will strongly interact. Such a prediction has been supported
by Stepanyuk{\it \ et al.} \cite{stepanyuk} who reported results
of first principle calculations of the exchange coupling between magnetic
impurities inside quantum corrals.

The physical properties of quantum corrals are affected by several factors,
which include confinement, quantum interference and many-body effects. So
far, interest has been focused on the static properties of quantum corrals,
leaving the dynamical behaviour of such systems almost unexplored.

In this work we have studied static and dynamical physical properties of
elliptical quantum corrals for various models and perturbations, where the
focalizing properties of the system have important consequences in the
physical behaviour.

We have considered two cases. The first one consisted of an isolated ellipse
describing a closed quantum system containing an arbitrary number of
non-interacting electrons up to a certain Fermi energy. The system was
perturbed by either local potentials located on the foci of the ellipse, or
a potential in the form of a barrier cutting the ellipse through its minor
axis. In the second case, two impurity spins interacting via a superexchange
term with the electrons in the ellipse were added to the foci of the system.
This many-body problem was treated numerically and static properties like
the spin gap, and dynamical responses, including real-time behaviour, where
analyzed.

\section{Non-interacting case}

Throughout this work we have adopted a simplified description of the system,
based on a spinless tight-binding model defined on a discretized square lattice within 
the 
boundaries of the ellipse, with eccentricity $\epsilon=0.6$ and hopping element 
$t^{\ast}$ between nearest neighbours. The Hamiltonian in this case is simple and 
reads $H_{el}=-t^{\ast}\sum_{\langle i, j \rangle} c^{\dagger}_i c_{j}$, where 
the 
sites $i$ and $j$ run inside the ellipse which consists of around 1000 sites or atoms. 
Here we have considered a constant zero local energy in each site.
As we will be analyzing
low-energy wave functions with wave lengths much larger that the real
interatomic distance, the underlying atomic distribution is irrelevant\cite{fiete1}. Our
calculations could have also relied on the exact solutions for a continuum
ellipse (Mathieu functions).

The Fermi
energy was set to coincide with its $23^{rd}$ eigenstate $|\psi_{23}\rangle $, which 
is similar to the 
state at the Fermi energy in the experimental setup of Ref. [1] (see Fig. 5(a)).
This configuration is not
crucial for the main conclusions and other eccentricities and reference
levels will have a similar behaviour, as long as the wave function bears an
appreciable weight at the foci of the ellipse. 

We have studied the time evolution of the wave function at the Fermi energy when, at time
$t=0$ a constant negative or positive potential $V_{1}$ is applied to focus 1:
$exp(-iHt)|\psi_{23}\rangle $, where $H=H_{el}+V_{1}(t)$. Figs. 1, 2 and 3 show the square
of the amplitude of such a wave function in foci 1 and 2 for $t\geq 0$, for three distinct
cases corresponding to $V_{1}=-1$ (Fig. 1), $V_{1}=1$ (Fig. 2) and $V_{1}=V_{2}=-1$ (Fig.
3). We notice that in the two first figures, a finite time $t_{0}$ elapses between the
onset of the perturbation in focus 1 and the observation of the response in focus 2, as
shown in the insets which is of the order of $t_{0}\sim 20\ \hslash /2t^{\ast 
}\sim10^{-14}$s, where we have used $t^{\ast }=1\ $eV (see below).

Another interesting point regards the oscillatory behaviour of the two squared amplitudes
for $t>t_{0}$. The occurrence of oscillations in the probability amplitude in the presence
of a time-independent potential (for $t>0$), as well as the overall ''see-saw" or
alternating behaviour between both foci, i.e., when the probability is high in one focus it
is low in the other and viceversa (except for the higher order oscillations in focus 1 for
negative $V_{1}$), are intimately related to quantum interference effects within the
ellipse. The period of these oscillations is of the order of the energy difference
corresponding to the main transitions involved, {\it i.e.} states 18, 23 and 29 for this
configuration.  Using the parameters of Refs. \cite{manoharan,aligia,weissmann} for the
case of the ellipse on the copper [111] surface, i.e. distance between foci $d=70$
$\mathring{A}$, total ellipse area $85.7$ nm$^{2}$ and $t^{\ast }=1\ $eV, we obtain 
periods 
of the order of $h /(E_{23}-E_{18})\simeq 5\times 10^{-14}$ s, where $E_{23}-E_{18}=0.0754\
t^{\ast }=75\ $meV (the transition between states 23 and 29 has a similar value).
We remark that this energy is one to two orders of magnitude larger
than the Kondo temperature in quantum dots. An animated version of these behaviours is
found in \cite{videos}. We notice that the amplitude of oscillation in focus 1 is larger
than that in focus 2 when $V_{1}$ is negative (attractive potential, Fig. 1), and smaller when
$V_{1}$ is positive (repulsive potential, Fig. 2).

\begin{figure}[ht]
\includegraphics[width=7cm,height=4cm]{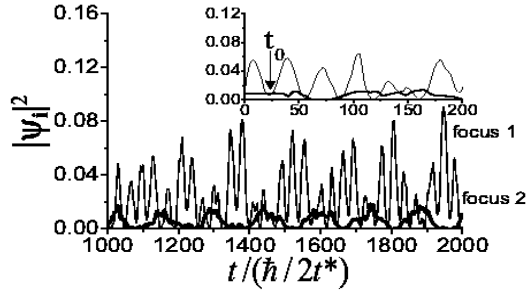}
\caption{Time dependence of the probability density in both foci for
a negative potential, $V_{1}=-1$, applied in focus 1 at time $t=0$.
Insets depict short times and $t_{0}$ is the time elapsed until the
perturbation reaches the unperturbed focus. For $t^{\ast }=1$ eV, typical
periods in this figure are around $5\times 10^{-14}$s.}
\end{figure}

\begin{figure}[ht]
\includegraphics[width=7cm,height=4cm]{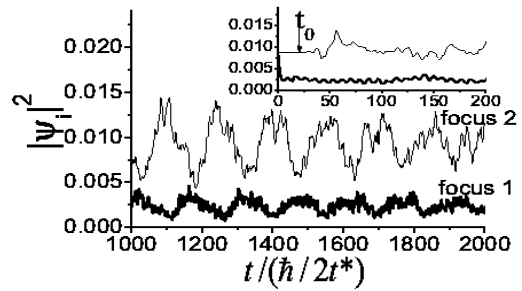}
\caption{Same as Fig. 1 but for a positive potential in focus 1, $V_{1}=1$.}
\end{figure}

\begin{figure}[ht]
\includegraphics[width=7cm,height=4cm]{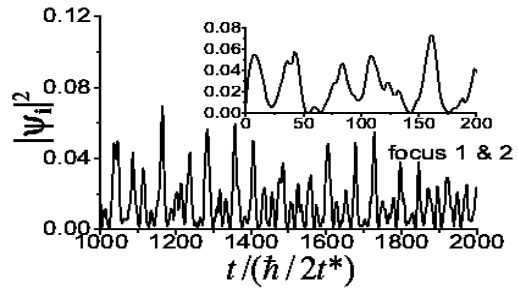}
\vspace{3mm}
\caption{Same as Fig. 1 but for a symmetric negative potential applied to foci 1 and 2, 
$V_{1}=V_{2}=-1$.}
\end{figure}

As a matter of corroboration, using the above set of parameters we obtain
for the Fermi velocity of the electrons involved in the transmission of the
information from one focus to the other the value $v_{F}=d/t_{0}=0.8\times
10^{8}$ cm/s (were we have considered the focal distance $d\simeq 70 \times 10^{-8}cm$ 
and $t_{0}\simeq 0.87 \times 10^{-14}s$), which is of the same order of 
magnitude as the one calculated in \cite{aligia}.

We have also calculated the delay time and Fermi velocity for parameters
corresponding to semiconductors, where, for an effective mass $m^{\ast
}=0.07m_{e}$ the energy difference of the main transition is around $1$eV
and $t_{0}=0.06\times 10^{-14}$ s, which leads to a Fermi velocity of around 
$10^{9}$cm/s and an electron density of $5.4\times 10^{13} cm^{-2}$, which is
somewhat higher than the typical carrier densities in semiconductor
heterostructures \cite{tan,ferry}.

A particularly interesting case was the response to an infinitely large
potential barrier along the minor axis applied at times $t\geq 0$, starting
from the same state as in the previous case (state 23). In Fig. 4(a) we see
the evolution which is quite periodic with a frequency corresponding to the
energy difference between the most relevant states involved (states 15, 16
and 25, 26 of the semi-ellipses, as shown in the approximate evolution where
only these states were included). We, thus, observe here that due to the
symmetry of the problem and the energy structure of the ellipse, only very
few states take part and the time evolution becomes simple and robust.

\begin{figure}[ht]
\includegraphics[width=7cm,height=7cm]{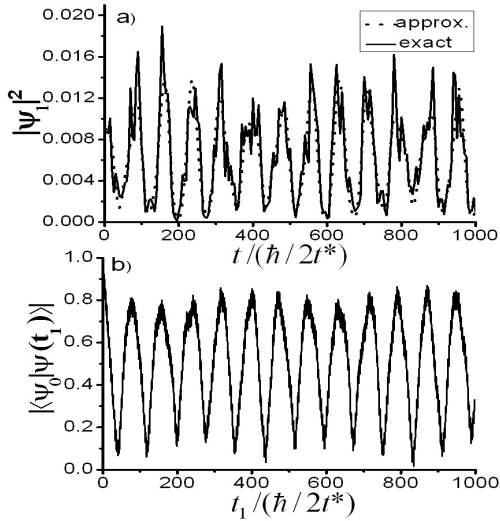}
\caption{a) Time dependence of the probability density in focus 1 for an
infinite potential barrier applied along the minor axis at time $t=0$. Solid
and dot lines correspond to the time evolution of the exact and aproximate
wave functions respectively. b) Modulus of the overlap between the
non-perturbed and final wave functions after the potential barrier had been
applied during an interval $t_1$ and then removed (note the different
parameters in the x-axis used both figures).}
\end{figure}

If, on the other hand, we leave the barrier potential on only for a finite time 
$t_{1}$ and then calculate the overlap of the original wave function to the
remaining one after having removed the perturbation, we find another
oscillatory behaviour between a nearly complete recovery of the original
wave function and a nearly completely orthogonal state (see Fig. 4(b)).

\section{Interacting case}

In addition to the one-particle behaviour studied in the previous section,
we have analyzed the more interesting case of the effect of the focalizing
properties of the ellipse on the interaction between two spins interacting
via a superexchange term with the electron spins in the ellipse (Fig. 5). The
Hamiltonian of the system reads:

\begin{equation}
H\!=\!H_{el} + J( \vec{S}_{1} \cdot \vec{\sigma}_{1} + \vec{S}_{2} \cdot \vec{\sigma}_{2}),
\end{equation}

where 

\begin{equation}
\vec{S}_i \cdot \vec{\sigma}_i \!=\! S_i^z \cdot \sigma_i^z + \frac{1}{2}(S_i^+ \cdot
\sigma_i^- + S_i^- \cdot \sigma_i^+),
\end{equation}

$\sigma_i^+\!=\!c_{i\uparrow}c_{i\downarrow}$, $\sigma_i^z\!=\!
(n_{i\uparrow}-n_{i\downarrow})/2$, with $n_{i\sigma}$ the number operator
and $c_{i\sigma}$ the destruction operator of an electron with spin $\sigma$
in focus $i$ of the ellipse. In the basis of eigenstates $|\alpha\rangle$ of
the ellipse, these local operators can be expanded as $c_{i\sigma}\!=\!\sum_{%
\alpha} \Psi_{\alpha i}c_{\alpha \sigma}$ where $c_{\alpha \sigma}$ and $%
\Psi_{\alpha i}$ are the destruction operator and amplitude in state $%
|\alpha\rangle$. $H_{el}$ is the Hamiltonian of the isolated ellipse
with infinite walls described in Sect. II.

In this basis the spin operators are expressed as: 
\begin{eqnarray}
\sigma_i^z&=&\frac{1}{2}\sum_{\alpha_1 \alpha_2} \Psi^*_{\alpha_1
i}\Psi_{\alpha_2 i} (c_{\alpha_1 \uparrow}^\dagger c_{\alpha_2 \uparrow} -
c_{\alpha_1 \downarrow}^\dagger c_{\alpha_2 \downarrow})  \nonumber \\
\sigma_i^+&=&\sum_{\alpha_1 \alpha_2} \Psi^*_{\alpha_1 i}\Psi_{\alpha_2 i}
c_{\alpha_1 \uparrow}^\dagger c_{\alpha_2 \downarrow}
\end{eqnarray}

\begin{figure}[ht]
\smallskip
\includegraphics[width=7cm,height=6cm]{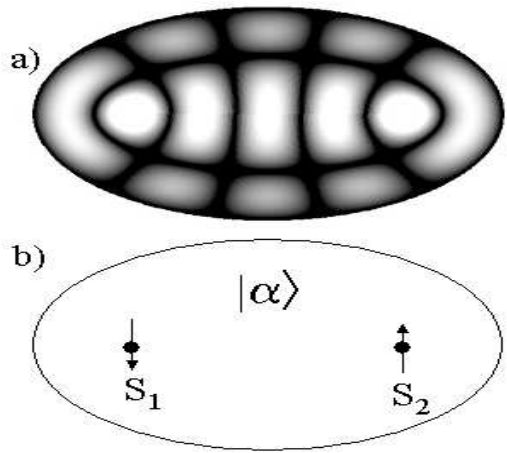}
\vspace{3mm}
\caption{a) $|\psi_{23}|^2$ for the non-interacting ellipse with eccentricity 
$\epsilon=0.6$. White and black represent high and low electron densities respectively.
b) Graphical representation of the system considered here where $|%
\protect\alpha\rangle$ represents an eigenstate of the non-perturbed ellipse.
$S_1$ and $S_2$ are the impurity spins at the foci of the ellipse which interact with the 
localized spins $\sigma_i$ via a superexchange interaction $J$.}
\label{fig5}
\end{figure}

\subsubsection{Static properties}

We have solved this Hamiltonian numerically with exact diagonalization for a
small number of levels and using the Lanczos technique for larger systems
and different fillings corresponding to an even (closed shell) or odd (open
shell) number of particles, $N$. The results for the spin gap $\Delta_s$ are
depicted in Fig. 6 where we find a different behaviour for even and odd
number of electrons in the ellipse. For the former case we observe a slow
increase of the spin gap as a function of the interaction parameter $J$
which is quadratic and linear for small and large $J$ respectively. For an
odd number of electrons the behaviour is linear from the beginning and the
gap is somewhat larger than for the even filling. One can understand this
behaviour by using perturbation theory. For the even case (closed shell) the
first correction to the energy stems from second-order non-degenerate terms
while for the odd filling (open shell), a first-order degenerate expansion
is required, leading to a linear dependence.

\begin{figure}[ht]
\includegraphics[width=8cm,height=6cm]{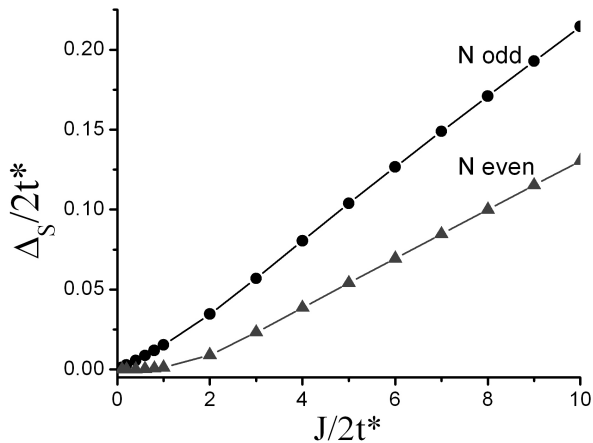}
\caption{Behaviour of spin gap $\Delta_s$ with $J$ for even (triangules) and
odd (circles) number of electrons in the ellipse. }
\end{figure}

For a realistic value \cite{chen} of the superexchange parameter for the Cu
ellipse, $J_{exp}=0.1$eV, we obtain $J/2t^{\ast }\simeq 0.05$ and a
corresponding gap of around $10^{-5}$ eV$\simeq 0.12$ K for even filling,
temperature under which the system is in a stable singlet state. For the
semiconductor ellipse the spin gap could be around 10 times larger although
we do not have any reliable estimation of $J$ in this case. Thus, in both cases, 
copper or semiconducting corrals, this
singlet state could be observed at sufficiently low temperatures.

One of the most important consequences of impurity spins in the presence of
an elliptical corral like the one considered here is their enhanced
correlation when situated at the foci of the system. For example, by
analyzing the spin-spin correlation function in the ground state $\langle
\psi_0| \vec{S}_1 \vec{S}_2 |\psi_0 \rangle$ we find that they form a
singlet or triplet for small $J$ and even or odd particles, respectively
(see Fig. 7(a)). A ferromagnetic interaction between localized spins for odd numbers 
of
electrons in the ellipse was also found in \cite{armandochiappe}. For larger
values of the interaction parameter both impurity spins decorrelate. The
results are weakly dependent on the number of levels considered, as also
shown here.

\begin{figure}[ht]
\smallskip
\includegraphics[width=7cm,height=7cm]{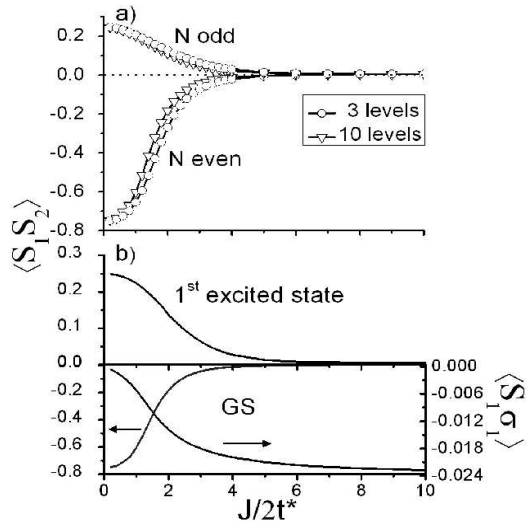}
\vspace{3mm}
\caption{a) Spin-spin correlation between localized spins in the ground
state for even and odd number of electrons and considering different number
of levels in the ellipse. b) Top panel: same as (a), $N$ even, for the first
excited state (having total spin $S=1$) showing the triplet character
between the localized spins in both foci for small $J$. Lower panel:
Correlation function between the localized spin $S_1$ and the spin of the
itinerant electron $\protect\sigma_1$  compared to the spin-spin 
correlation
between localized spins for even $N$ (for the ground state in all cases). For small $J$ the 
localized spins are
entangled in a singlet state with no correlation with the electrons. For
large $J$, these spins decouple while enhancing the local singlet
interaction with the itinerant electron.}
\end{figure}

It is also interesting to analyze the on-site correlation between the
localized spin $S_1$ and the spin of the electron localized immediately
above it, $\sigma_1$. In Fig. 7(b) we show how this interaction, which is
antiferromgnetic, strengthens by increasing $J$, while $S_1$ decorrelates
from $S_2$. In addition, as $J$ increases, the electrons in the ellipse
become more localized near the impurity in order to take advantage of the
magnetic interaction and the spin-spin correlation should grow towards the
singlet value -0.75. We find a lower absolute value for large $J$ and this
is due to the finiteness of the system. In this limit, many more extended
states are required to form the localized state (see Eq. 3).

\subsubsection{Dynamical behaviour}

We have calculated the spin response function to a spin excitation performed
in focus 1\cite{mahan} 
\begin{equation}
A(\omega)=-\frac{1}{\pi} \lim_{\eta\to 0^+} Im G(\omega+i\eta +E_0)
\end{equation}
where 
\begin{equation}
G(z)=\langle \psi_0 | S_1^z (z-H)^{-1} S_1^z |\psi_0 \rangle .
\end{equation}

\begin{figure}[ht]
\smallskip
\includegraphics[width=7cm,height=7cm]{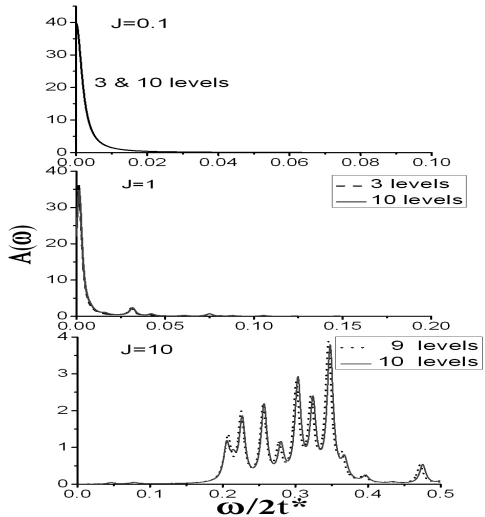}
\vspace{3mm}
\caption{Spin spectral function $A(\protect\omega)$ for an even number of
particles, three values of $J$ and different number of levels in the ellipse
showing a negligible ''finite size" effect. For small values of $J$ the spin
gap is small and the first excitation is mainly a triplet state. See also
Fig.7(b), top panel.}
\end{figure}

\begin{figure}[ht]
\vspace{-10mm}
\includegraphics[width=7cm,height=7cm]{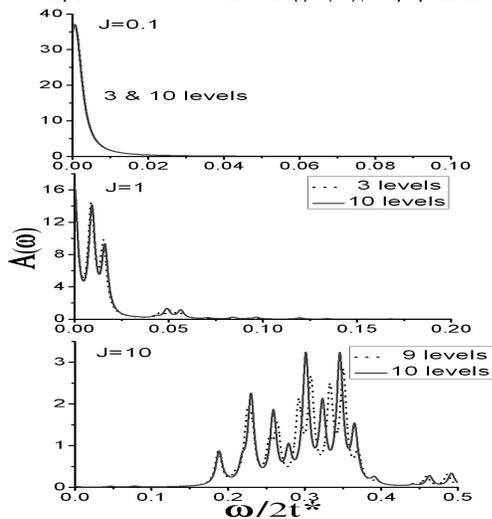}
\vspace{3mm}
\caption{Same as Fig. 8 for odd $N$}
\label{fig9}
\end{figure}

\begin{figure}[ht]
\smallskip
\includegraphics[width=7cm,height=7cm]{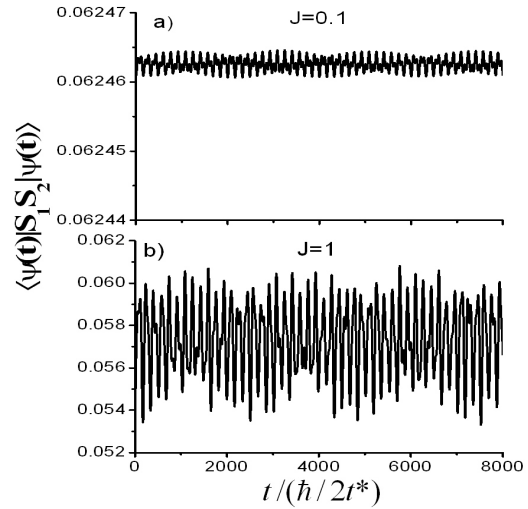} 
\vspace{3mm}
\caption{Time dependence of the spin-spin correlation function between
localized spins for the perturbed wave function $|\protect\psi(t)
\rangle=U(t) S_1^z |\protect\psi_0 \rangle $ for small and moderate
interactions. For small $J$ the correlation is nearly a triplet as seen in
previous figures.}
\label{fig10}
\end{figure}

In Figs. 8 and 9 we show the results for different values of $J$ and for
even and odd fillings, respectively. For even $N$ we find that the
excitations for small $J$ consist mainly of one peak located at the spin
gap, which is small. In this case both localized spins form a very well
defined singlet, whereas the first spin excitation has a triplet character
as can be seen in Figs. 7(b) (top panel) and 10(a) (see below). For larger
values of $J$ the spectral weight of this peak diminishes, other states get
involved and the spectral function is more complicated (see also larger
values of $J$ in Fig. 7(b)). For odd $N$ and small $J$ both localized spins
form a well defined triplet in the ground state and the first spin
excitation corresponds to a spin flip of an electron spin, leaving $S_1$ and 
$S_2$ still entangled in a triplet state. In both cases a negligible ''size"
dependence with the number of levels considered is found.


The fact that, for small $J$ and an even number of particles, the first
excitation is nearly purely a triplet can also be seen as a robust
excitation which varies very slowly in time (see Fig. 10). Here we depict $%
\langle \psi(t)| \vec{S}_1 \vec{S}_2|\psi(t) \rangle $ where $|\psi(t)
\rangle=U(t)S_1^z|\psi_0 \rangle$ and $U(t)$ the time-evolution operator and 
$|\psi_0 \rangle$ is the interacting ground state. For $J=0.1$ we find a
nearly constant behaviour, scrumbling up with a larger amount of eigenstates
for larger values of $J$. Note that the factor 1/4 which arises when
comparing this correlation to a pure triplet behaviour stems from the
perturbation operator $S_1^z$ applied to the ground state (the function $%
|\psi(t) \rangle$ is not normalized).

\section{Conclusions}

We have studied static and dynamical properties of hard-wall elliptical quantum
corrals subject to different kinds of perturbations. On one hand we
considered non-interacting systems and applied localized perturbation
potentials in the foci and along the minor axis of the ellipse, finding
robust oscillatory behaviours which are a consequence of the small amount of
levels involved in the excitations. A finite elapse was found in the case
that the potential was applied to one focus until the perturbation reaches
the unperturbed one. On the other hand we studied the interacting case where
two localized spins at the foci of the ellipse interacted
antiferromagnetically with the itinerant electrons. In this case we found a
different behaviour for even and odd electrons which shows up in the spin
gap, the character of the ground state and the excitations. For the even
particle number case and small interaction $J$, both localized spins are entangled
in a (quasi)-singlet state, forming a (quasi)-triplet in the first excitated
state, which is the main spin excitation. These states are very robust,
however, the spin gap is quadratically small with $J$. An interesting
feature can be seen which resembles the RKKY-Kondo like transition occurring
in the two-impurity system when increasing the Kondo interaction $J$: While
the localized spins form a singlet state for small interaction, this
coupling decreases for larger $J$ giving rise to an on-site singlet
correlation between the spin and the itinerant electron.

In this work we have shown that the focalizing properties of elliptic
corrals lead to an interesting behaviour between localized spins. For small $%
J$ and even number of electrons in the ellipse, the system can be modelled
by two spins interacting via an effective antiferromagnetic interaction
parameter $H_{eff}=J_{eff}\vec{S}_1 . \vec{S}_2$, where $J_{eff}=\Delta_s$.
For odd number of electrons the interaction is ferromagnetic in the ground
state. The character of this interaction can be controlled by changing the
chemical potential of the system.

For small to moderate values of $J$ ($ J \lesssim 1$) the main results synthesized 
above still
hold in the case of more realistic models of quantum ellipses which include tunneling of
the electrons in open corrals and inelastic processes with bulk electrons. In 
this
parameter range the broadening of the relevant energy levels is smaller than their
separation\cite{nos2,fiete1,calderon,crampin}. When larger interactions are included,
higher levels which are more hybridized take part, and models including these 
processes should be considered.\newline \newline

K.H. acknowledges support from the  Guggenheim Foundation. We thank CONICET, PIP 5254,
PICT 03-12742 and 03-13829 of ANPCyT of Argentina for support. Partial
financial support from CNPq, Fundo PROSUL, FAPERJ, and Millenium Institute
of Nantechnology/MCT of Brazil is gratefully acknowledged. We are also
grateful to A. Aligia for a critical reading of the manuscript.

\end{document}